\documentclass[pra,aps,twocolumn,showpacs,superscriptaddress]{revtex4}

\usepackage{amsmath}
\usepackage{bm}
\usepackage{graphicx}

\begin{document}

\title{Many-body dynamics of a Bose--Einstein condensate collapsing by
quantum tunneling
}

\author{Hiroki Saito}
\affiliation{Department of Engineering Science, University of
Electro-Communications, Tokyo 182-8585, Japan}

\date{\today}

\begin{abstract}
The dynamics of a Bose--Einstein condensate of atoms having attractive
interactions is studied using quantum many-body simulations.
The collapse of the condensate by quantum tunneling is numerically
demonstrated and the tunneling rate is calculated.
The correlation properties of the quantum many-body state are
investigated.
\end{abstract}


\maketitle

\section{Introduction}

Attractive interactions between particles destabilize many-body systems
against collapse or aggregation.
Examples include gravitational collapse of a star and nucleation of drops
from supercooled water vapor.
Owing to this kind of instability, a dilute Bose gas with attractive
interactions cannot form a stable Bose--Einstein condensate (BEC) in an
infinite homogeneous system~\cite{Stoof94}.
In a finite system, by contrast, quantum pressure arising from zero-point
energies can sustain an attractive force and a metastable BEC can
form~\cite{Ruprecht}.
Such a metastable BEC with attractive interactions was first realized with
trapped $^7{\rm Li}$ atoms~\cite{Bradley95}.
A metastable BEC becomes unstable against collapse when the number of
atoms or the scattering length exceeds a critical value~\cite{Roberts}.
As a result, the number of $^7{\rm Li}$ atoms in the BEC was
limited~\cite{Bradley97}.
When atoms are continuously replenished in the BEC, repeated collapse and
growth occur ~\cite{Kagan98,Sackett98,Sackett99,Gerton}.
The collapse can be investigated in a controlled manner by tuning the
interatomic interaction via a Feshbach resonance.
The collapsing dynamics of an $^{85}{\rm Rb}$ BEC has been studied using
this technique.
The interaction energy is converted into kinetic energy during the
collapse, and atomic bursts and jets are
produced~\cite{Cornish,Donley,Altin}.
A Feshbach resonance has also been used to study the amplification of
local instabilities~\cite{Chin}.
The dynamics of collapsing and exploding BECs have been theoretically
investigated by many
researchers~\cite{Pitaevskii,Kagan97,Elef,Saito,Duine,Yurovsky,Santos,Adhikari,Milstein,Savage,Calzetta,Metens,Bao,Wuster}.

A trapped metastable BEC with an attractive interaction can collapse
by macroscopic quantum tunneling.
The tunneling rate can be estimated from an overlap
integral~\cite{Kagan96,Shuryak} or a path integral over the
semiclassical trajectory~\cite{Stoof,Leggett,Ueda,Huepe}.
However, the dynamics of a BEC collapsing by quantum tunneling have not
been studied.
The mean-field approximation, widely used to study BECs, cannot be applied
in this situation, because it neglects many-body quantum fluctuations and
the metastable state never collapses if the Gross--Pitaevskii equation is
used.

In the present paper, direct quantum many-body simulations are preformed
to investigate the dynamics of an attractive BEC in a metastable state,
and collective quantum many-body collapse by quantum tunneling is
demonstrated.
Owing to the limited computational resource, the system is restricted to a
few dozen atoms.
The tunneling decay rate is determined for a metastable BEC, and its
dependence on the scattering length and on the number of atoms is obtained.
It is shown that the quantum state develops into a superposition between
metastable and collapsing states.

The paper is organized as follows.
Section~\ref{s:formulation} formulates the problem and the numerical
methods.
Section~\ref{s:results} presents the numerical results.
Section~\ref{s:conc} ends with the conclusions.

\section{Formulation and numerical methods}
\label{s:formulation}

Consider a system of bosonic atoms of mass $m$ confined in an external
potential $V(\bm{r})$.
The interaction between atoms is approximated as a contact potential with
an $s$-wave scattering length $a$, which is negative for attractive
interactions.
The Hamiltonian for the system is
\begin{eqnarray} \label{H}
\hat H & = & \int d\bm{r} \hat\psi^\dagger(\bm{r}) \left[
-\frac{\hbar^2}{2m} \nabla^2 + V(\bm{r}) \right] \hat\psi(\bm{r})
\nonumber \\
& & + \frac{2\pi\hbar^2 a}{m}
\int d\bm{r} \hat\psi^{\dagger 2}(\bm{r}) \hat\psi^2(\bm{r}),
\end{eqnarray}
where the field operator $\hat\psi(\bm{r})$ annihilates an atom located at
$\bm{r}$.
For simplicity, an isotropic harmonic trap of frequency $\omega$ is used,
\begin{equation} \label{V}
V(\bm{r}) = \frac{1}{2} m \omega^2 r^2,
\end{equation}
where $r^2 = x^2 + y^2 + z^2$.

Although the quantum many-body dynamics will be studied, it is helpful to
introduce a mean-field analysis.
The Gross--Pitaevskii (GP) equation for the system described by
Eq.~(\ref{H}) has the form,
\begin{equation} \label{GP}
i\hbar \frac{\partial \psi}{\partial t} = \left[ -\frac{\hbar^2}{2m}
\nabla^2 + V(\bm{r}) \right] \psi + \frac{4\pi\hbar^2 a (N - 1)}{m}
|\psi|^2 \psi,
\end{equation}
where $\psi(\bm{r}, t)$ is the macroscopic wave function normalized as
$\int |\psi|^2 d\bm{r} = 1$ and $N$ is the number of atoms.
A metastable stationary wave function $\psi_{\rm ms}(r)$
exists~\cite{Ruprecht} for interaction strengths $|g|$ smaller than a
critical value $|g_{\rm c}|$,
\begin{equation} \label{gc}
g \equiv 4\pi a (N - 1) \sqrt{\frac{m\omega}{\hbar}} > g_{\rm c} \simeq
-7.225,
\end{equation}
for an isotropic harmonic trap.
When $|g|$ exceeds $|g_{\rm c}|$, any wave function eventually collapses.
On the other hand, when $|g| < |g_{\rm c}|$, the excitation frequencies
above state $\psi_{\rm ms}$ are real and positive, and therefore
$\psi_{\rm ms}$ is stable as long as it evolves according Eq.~(\ref{GP}).
Numerically, a metastable wave function $\psi_{\rm ms}$ can be computed
using the imaginary-time propagation method~\cite{Dalfovo}, where $i$ on
the left-hand side of Eq.~(\ref{GP}) is replaced by $-1$.

Returning to a quantum many-body analysis, the number $M$ of basis
functions must be restricted in the  numerical calculations.
Cut off the field operator as
\begin{equation} \label{psi}
\hat\psi(\bm{r}) = \sum_{j = 1}^M \hat a_j \phi_j(\bm{r}),
\end{equation}
where $\hat a_j$ is the bosonic operator satisfying $[\hat a_j, \hat
a_{j'}^\dagger] = \delta_{jj'}$.
Using this basis, a many-body state $|\psi \rangle$ can be expanded by
Fock states as
\begin{equation} \label{ket}
|\psi \rangle = \sum_{n_1, \cdots, n_M} c_{n_1, \cdots, n_M}
|n_1, \cdots, n_M \rangle,
\end{equation}
where
\begin{equation} \label{Fock}
|n_1, \cdots, n_M \rangle = \prod_{j = 1}^M
\frac{(\hat a_j^\dagger)^{n_j}}{\sqrt{n_j!}} |0\rangle
\end{equation}
with $|0\rangle$ being the vacuum state.
The summation in Eq.~(\ref{ket}) is over non-negative integers satisfying
\begin{equation} \label{sumN}
\sum_{j = 1}^M n_j = N.
\end{equation}

Choose the basis functions $\phi_j(\bm{r})$ in Eq.~(\ref{psi}) as
follows.
To express the quantum states before collapse, a reasonable choice of
basis function is the metastable state of the GP equation:
\begin{equation}
\phi_1(\bm{r}) = \psi_{\rm ms}(r).
\end{equation}
To simulate the collapsing dynamics, in which the state shrinks
isotropically, consider functions scaled from $\psi_{\rm ms}(r)$ according
to
\begin{equation}
f_j(r) = \psi_{\rm ms}(r / \alpha^j),
\end{equation}
where $j = 2, \cdots, M$ with a constant $0 < \alpha < 1$.
In the numerical calculations presented next, the value of $\alpha$ is
taken to lie in the range $0.6$-$0.7$.
Using Gram--Schmidt orthonormalization, the other $M - 1$ basis
functions are
\begin{equation}
\phi_j(\bm{r}) = {\cal N}_j \left[ f_j(r) - \sum_{n=0}^{j-1}
\int \phi_n^*(r) f_j(r) d\bm{r} \right],
\end{equation}
where ${\cal N}_j$ is a normalization factor.
Since the collapse occurs isotropically for the potential of Eq.~(\ref{V}),
anisotropic quantum fluctuations can be neglected and the basis functions
$\phi_j(r)$ can be taken to be isotropic.

Substitution of Eq.~(\ref{psi}) into Eq.~(\ref{H}) gives
\begin{equation}
\hat H = \sum_{j_1, j_2}^M K_{j_1 j_2} \hat a_{j_1}^\dagger \hat a_{j_2}
+ \sum_{j_1, j_2, j_3, j_4}^M I_{j_1 j_2 j_3 j_4} \hat a_{j_1}^\dagger
a_{j_2}^\dagger \hat a_{j_3} \hat a_{j_4},
\end{equation}
where
\begin{equation}
K_{j_1 j_2} = \int d\bm{r} \phi_{j_1}^*(r) \left[ -\frac{\hbar^2}{2m}
\nabla^2 + V(\bm{r}) \right] \phi_{j_2}(r),
\end{equation}
and
\begin{equation}
I_{j_1 j_2 j_3 j_4} = \frac{2\pi\hbar^2 a}{m} \int d\bm{r}
\phi_{j_1}^*(r) \phi_{j_2}^*(r) \phi_{j_3}(r) \phi_{j_4}(r).
\end{equation}
The number of Fock states in Eq.~(\ref{Fock}) that satisfy
Eq.~(\ref{sumN}) is
\begin{equation} \label{num}
n_{\rm F}(N, M) = \frac{(N + M - 1)!}{N! (M - 1)!}.
\end{equation}
In terms of the Fock basis, the quantum state in Eq.~(\ref{ket}) is a
vector having $n_{\rm F}$ components and the Hamiltonian is an $n_{\rm F}
\times n_{\rm F}$ matrix.
The many-body Schr\"odinger equation,
\begin{equation} \label{Sch}
i \hbar \frac{\partial}{\partial t} |\psi(t) \rangle
= \hat H |\psi(t) \rangle,
\end{equation}
then becomes simultaneous differential equations for the vector of
$n_{\rm	F}$ components, $c_{n_1, \cdots, n_M}(t)$ in Eq.~(\ref{ket}),
which are time-integrated using a fourth-order Runge--Kutta method.
In the numerical calculations, it is convenient to create a dictionary
that provides $\{ n_j \}$ given the vector index $\ell = 1, \cdots,
n_{\rm F}$.
To obtain the vector index $\ell$ from $\{ n_j \}$, one can use
\begin{equation}
\ell = 1 + \sum_{k = 1}^{M - 1} \left( \begin{array}{c} N - \sum_{j=1}^k
n_j + M - k - 1 \\ M - k \end{array} \right).
\end{equation}

The initial state is taken to be $| N, 0, 0, \cdots, 0 \rangle$, i.e., all
the atoms occupy wave function $\psi_{\rm ms}(r)$, which is a good
starting point for the metastable many-body state.
To obtain the decay rate of the metastable state by quantum tunneling, we
must identify and eliminate the ``collapsed state,'' which is the portion
of the quantum state that has collapsed and never returns to the original
state.
Since the interaction energy is converted to kinetic energy during the
collapse, it is reasonable to assume that the collapsed state consists of
Fock states with large kinetic energies.
Therefore a non-Hermitian term is added to the Hamiltonian, such as
\begin{equation} \label{loss}
-i L \sum | n_1, \cdots, n_M \rangle \langle n_1, \cdots, n_M |,
\end{equation}
where the summation is taken over states satisfying
\begin{equation}
\sum_{j=1}^M n_j K_{jj} > E_{\rm threshold}.
\end{equation}
Take $L \sim 10 \hbar \omega$ and $E_{\rm threshold}$ to be several times
larger than $N K_{11}$.
The results hardly depend on these values, implying that the collapsed
part of the quantum state is effectively eliminated by this method.
If the collapsed portion is left untreated, it bounces back to the
original state because of the restricted number of basis functions,
and the correct decay rate cannot be obtained.

\section{Numerical results}
\label{s:results}

To demonstrate many-body collapse by quantum tunneling, Eq.~(\ref{Sch}) is
numerically solved for a value of the interaction coefficient $g$ near
its critical value $g_{\rm c}$.
Since the critical value for the collapse is ambiguous in a quantum
many-body analysis, the mean-field value $g_{\rm c}$ in Eq.~(\ref{gc}) is
adopted as the critical value.

\begin{figure}[tbp]
\includegraphics[width=8cm]{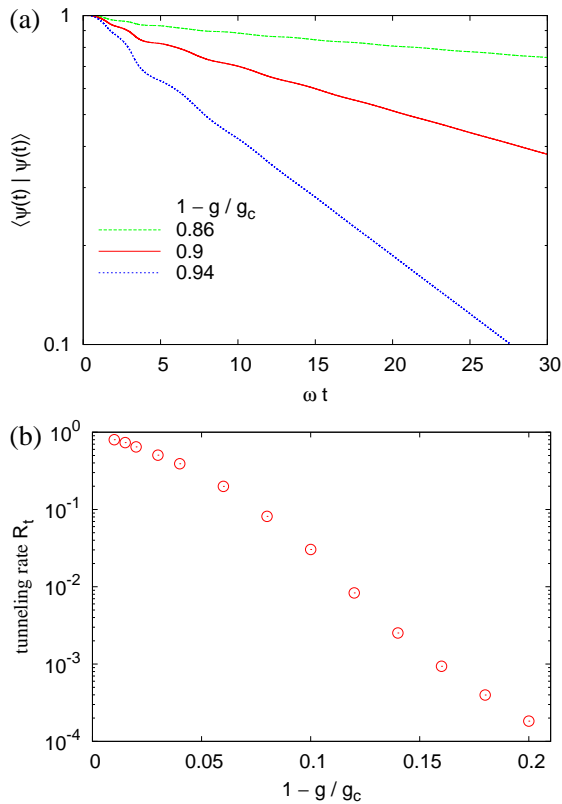}
\caption{
(color online) (a) Time evolution of $\langle \psi(t) | \psi(t) \rangle$
for $1 - g / g_{\rm c} = 0.86$ (dashed line), $0.9$ (solid line), and
$0.94$ (dotted line).
(b) Tunneling rate $R_{\rm t}$ as a function of $1 - g / g_{\rm c}$, where
$R_t$ is obtained from the slope of the lines in (a).
The number of atoms is $N = 32$ and the number of mode functions is $M =
6$.
}
\label{f:decay}
\end{figure}
Figure~\ref{f:decay}(a) plots the time evolution of the norm of the
many-body state in Eq.~(\ref{ket}),
\begin{equation}
\langle \psi(t) | \psi(t) \rangle = \sum_{n_1, \cdots, n_M}
|c_{n_1, \cdots, n_M}(t)|^2,
\end{equation}
which is unity at $t = 0$.
For $\omega t \lesssim 1$, $\langle \psi(t) | \psi(t) \rangle$ remains at
unity, because it takes a time for the condensate to collapse, as observed
in experiments~\cite{Donley}.
After $\omega t \sim 1$, a portion of the quantum state undergoes
tunneling and collapse, and is removed by the non-Hermitian term in
Eq.~(\ref{loss}), resulting in the decay of $\langle \psi(t) | \psi(t)
\rangle$ seen in Fig.~\ref{f:decay}(a).
For $\omega t \gtrsim 10$, the fluctuations in the curves in
Fig.~\ref{f:decay}(a) attenuate and $\log \langle \psi(t) | \psi(t)
\rangle$ decreases linearly in time, indicating that the correct tunneling
decay rates are obtained no matter how the initial states are chosen.
Define the tunneling rate $R_{\rm t}$ from
\begin{equation}
\langle \psi(t) | \psi(t) \rangle \sim e^{-R_{\rm t} t},
\end{equation}
obtained from the slope of the lines in Fig.~\ref{f:decay}(a).
Figure~\ref{f:decay}(b) shows the tunneling rate $R_{\rm t}$ as a function
of $1 - g / g_{\rm c}$.
The tunneling rate $R_{\rm t}$ decreases exponentially with increasing
value of $1 - g / g_{\rm c}$.
However, Fig.~\ref{f:decay}(b) does not prove whether the tunneling rate
obeys $\log R_{\rm t} \propto -(1 - g / g_{\rm c})$~\cite{Shuryak} or  
$\log R_{\rm t} \propto -(1 - g / g_{\rm c})^{5/4}$~\cite{Leggett,Huepe}.
In these previous studies, the tunneling rate was calculated by the
semiclassical approximation valid for a large number of atoms $N$,
whereas the present study only uses a few dozen atoms.

\begin{figure}[tbp]
\includegraphics[width=8cm]{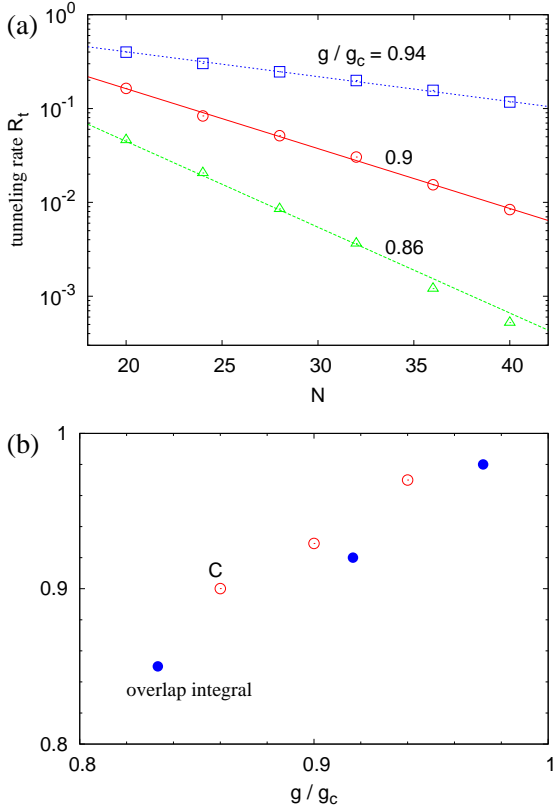}
\caption{
(color online) (a) Tunneling rate $R_{\rm t}$ as a function of the number
of atoms $N$, where $g$ is fixed at value $0.86 g_{\rm c}$ (triangles),
$0.9 g_{\rm c}$ (circles), and $0.94 g_{\rm c}$ (squares).
(b) The open circles plot $C = e^{s / 2}$, where $s$ is the slope of the
lines in (a).
The filled circles plot the value of the overlap integral
$\left| \int \psi_{\rm outside}^*(r) \psi_{\rm ms}(r) d\bm{r} \right|$
from Ref.~\cite{Shuryak}.
The number of mode functions is $M = 6$.
}
\label{f:ndep}
\end{figure}
Figure~\ref{f:ndep}(a) plots the dependence of the tunneling rate
$R_{\rm t}$ on $N$, while the value of $g$ in Eq.~(\ref{gc}) is fixed at
either $0.86 g_{\rm c}$, $0.9 g_{\rm c}$, or $0.94 g_{\rm c}$.
The tunneling rate $R_{\rm t}$ exponentially decreases with increasing
$N$.
The slopes of the lines in Fig.~\ref{f:ndep}(a) are defined as $s$, i.e.,
\begin{equation} \label{ndep}
R_{\rm t} \propto e^{s N}.
\end{equation}
The tunneling rate can be estimated by the overlap integral between the
mean-field wave functions~\cite{Shuryak},
\begin{equation} \label{overlap}
R_{\rm t} \sim \left| \int \psi_{\rm outside}^*(r) \psi_{\rm ms}(r)
d\bm{r} \right|^{2N},
\end{equation}
where $\psi_{\rm ms}(r)$ is the metastable wave function and
$\psi_{\rm outside}(r)$ is a wave function outside the barrier against
collapse.
It follows from Eqs.~(\ref{ndep}) and (\ref{overlap}) that $e^{s / 2}$
corresponds to the mean-field overlap integral $\left| \int
\psi_{\rm outside}^*(r) \psi_{\rm ms}(r) d\bm{r} \right|$.
The values of $e^{s / 2}$ and the mean-field overlap integral are plotted
in Fig.~\ref{f:ndep}(b).
They are seen to be in reasonable agreement.
The open circles in Fig.~\ref{f:ndep}(b) lie slightly above the filled
circles, probably because there are many paths to collapse other than
$\psi_{\rm outside}$.

Before analyzing the properties of the many-body quantum state, consider
it qualitatively.
Schematically, the quantum state may be written as
\begin{equation} \label{schematic}
|\psi_{\rm metastable}\rangle + |\psi_{\rm collapsing}\rangle
+ |\psi_{\rm collapsed}\rangle,
\end{equation}
where $|\psi_{\rm metastable}\rangle$ is a nearly pure condensate having a
Gaussian shape, $|\psi_{\rm collapsing}\rangle$ is a state in which the
central density is increasing, and $|\psi_{\rm collapsed}\rangle$ is the
state in which the collapse has advanced and the central density has
become extremely large or an explosion has occurred.
In the present simulation, $|\psi_{\rm collapsed}\rangle$ is removed by
Eq.~(\ref{loss}), because it rapidly decoheres from
$|\psi_{\rm metastable}\rangle$ and $|\psi_{\rm collapsing}\rangle$ owing
to the fragility of the macroscopic superposition.
Thus the quantum state obtained in the present simulation is written as
\begin{equation} \label{schematic2}
|\psi \rangle \sim |\psi_{\rm metastable}\rangle
+ |\psi_{\rm collapsing}\rangle.
\end{equation}
Bear in mind that an expectation value is taken with respect to
Eq.~(\ref{schematic2}) in the following discussion, not with respect to
Eq.~(\ref{schematic}).

\begin{figure}[tbp]
\includegraphics[width=8cm]{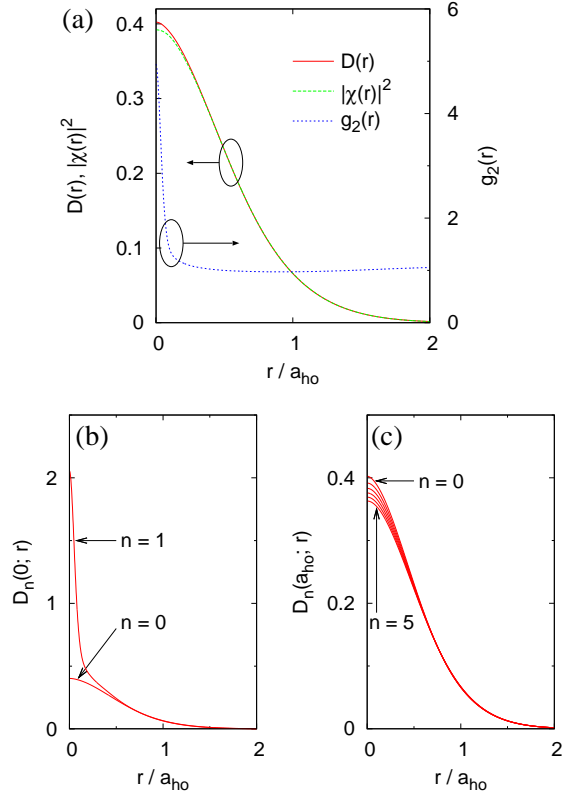}
\caption{
(color online) (a) Density distribution $D(r)$ from Eq.~(\ref{Dr}),
density $|\chi(r)|^2$ of the condensate wave function from Eq.~(\ref{chi}),
and the second-order correlation function $g_2(r)$ from Eq.~(\ref{g2}).
(b)-(c) Conditional density distribution $D_n(r_{\rm d}; r)$ from
Eq.~(\ref{Dn}) for (a) $r_{\rm d} = 0$ and (b) $r_{\rm d} = a_{\rm ho}$.
In (c), $n = 0, 1, \cdots, 5$ from top to bottom.
In (a)-(c), the parameters are $g / g_{\rm c} = 0.8$, $N = 32$, and
$M = 6$. 
The state $|\psi(t) \rangle$ at $\omega t = 30$ is used.
}
\label{f:detect}
\end{figure}
Now consider the density and correlation properties of a quantum
many-body state $|\psi \rangle$.
The solid curve in Fig.~\ref{f:detect}(a) plots the expectation value of
the atomic density,
\begin{equation} \label{Dr}
D(r) = \frac{\langle \psi | \hat \psi^\dagger(r) \hat \psi(r) |
\psi \rangle}{\langle \psi | \psi \rangle}.
\end{equation}
The condensate wave function $\chi(r)$ is obtained by diagonalizing the
single-particle density matrix,
\begin{equation}
\left( \begin{array}{ccc}
\langle \hat a_1^\dagger \hat a_1 \rangle & \cdots & 
\langle \hat a_1^\dagger \hat a_M \rangle \\
\vdots & \ddots & \vdots \\
\langle \hat a_M^\dagger \hat a_1 \rangle & \cdots & 
\langle \hat a_M^\dagger \hat a_M \rangle \end{array} \right).
\end{equation}
Using the normalized eigenvector $\bm{v}$ of this matrix having the
largest eigenvalue, the condensate wave function $\chi(r)$ is written as
\begin{equation} \label{chi}
\chi(r) = \sum_{j=1}^M v_j \phi_j(r).
\end{equation}
The density $|\chi(r)|^2$ is shown as the dashed curve in
Fig.~\ref{f:detect}(a), where the condensate fraction is 0.997.
For $r / a_{\rm ho} \lesssim 0.2$, the density $D(r)$ deviates from the
condensate density $|\chi(r)|^2$.
The dotted curve in Fig.~\ref{f:detect}(a) shows the second-order
correlation function
\begin{equation} \label{g2}
g_2(r) = \frac{1}{D^2(r)}
\frac{\langle \psi | \hat \psi^{\dagger 2}(r) \hat \psi^2(r) |
\psi \rangle}{\langle \psi | \psi \rangle},
\end{equation}
which has a sharp peak at the center, where the density fluctuation is
large.

The deviation of $D(r)$ from $|\chi(r)|^2$ and the large peak in $g_2(r)$
at the center of the BEC in Fig.~\ref{f:detect}(a) arise because the
quantum state has the form of Eq.~(\ref{schematic2}).
To verify that, the conditional density distribution after $n$ atoms are
detected at $r = r_{\rm d}$,
\begin{equation} \label{Dn}
D_n(r_{\rm d}; r) = {\cal N} \langle \hat\psi^{\dagger n}(r_{\rm d})
\hat\psi^\dagger(r) \hat\psi(r) \hat\psi^n(r_{\rm d}) \rangle,
\end{equation}
is calculated, where ${\cal N}$ is the factor normalizing
$\int D_n(r_{\rm d}; r) d\bm{r} = 1$.
For example, for a pure condensate, Eq.~(\ref{Dn}) is independent of $n$
and $r_{\rm d}$: $D_n(r_{\rm d}; r) = |\chi(r)|^2$.
Figures~\ref{f:detect}(a) and \ref{f:detect}(b) show $D_n(r_{\rm d}; r)$
for $r_{\rm d} = 0$ and $r_{\rm d} = a_{\rm ho} \equiv
[\hbar / (m \omega)]^{1/2}$.
If an atom is detected at the center of the trap $(r_{\rm d} = 0)$, the
state reduction enhances the second term in Eq.~(\ref{schematic2}),
resulting in an increase in the central density, as seen in
Fig.~\ref{f:detect}(a).
If an atom is detected at the periphery of the cloud, on the other hand, 
the second term in Eq.~(\ref{schematic2}) reduces, and the density
distribution changes as shown in Fig.~\ref{f:detect}(b).
It can thus be concluded that the quantum many-body state evolves to a
superposition between a metastable state with a Gaussian-like density
distribution and the small fraction of the collective collapsing state
having a  sharp central density.

For $1 < n < N$, the distribution $D_n(r_{\rm d} = 0; r)$ has a sharp
central peak similar to the curve for $n = 1$ in Fig.~\ref{f:detect}(a).
This fact indicates that the collective collapse dominates the collapsing
dynamics, in which $\sim N$ atoms participate in the collapse.
If the sharp central peak in $D_n(r_{\rm d} = 0; r)$ vanished for $n
\gtrsim N_1$, it would be regarded as a partial collapse~\cite{Kagan96},
in which a cluster of $N_1$ atoms collapses and $N - N_1$ atoms remain in
the Gaussian-like wave function.
However, that does not happen in the present simulation.

\section{Conclusions}
\label{s:conc}

The quantum many-body dynamics have been investigated for a BEC with
attractive interactions.
A numerical method has been developed that is suitable for simulating the
collapsing dynamics.
The method was applied to a system of a few dozen atoms to demonstrate the
collapse of a BEC by quantum tunneling.
This result is the first numerical simulation of a quantum many-body
system collapsing upon itself.
Due to the quantum tunneling, the uncollapsed component decays
exponentially (Fig.~\ref{f:decay}(a)), so that the tunneling rate can be
obtained.
The tunneling rate decreases exponentially with an increase in the value
of $1 - g / g_{\rm c}$ (Fig.~\ref{f:decay}(b)) and of $N$
(Fig.~\ref{f:ndep}(a)).
The tunneling rate is in reasonable agreement with the overlap integral
computed between the mean-field wave functions before and after the
tunneling (Fig.~\ref{f:ndep}(b)).
The quantum many-body state develops into a macroscopic superposition
between the uncollapsed and collapsing states (Fig.~\ref{f:detect}).

If the number of mode functions can be increased, then the exploding
dynamics can also be studied, revealing the quantum many-body properties
of the burst atoms.

\begin{acknowledgments}
This work was supported by a Grant-in-Aid for Scientific
Research (Nos.\ 23540464 and 25103007) from the Ministry of Education,
Culture, Sports, Science and Technology of Japan.
\end{acknowledgments}

\end{document}